# Towards Secure SPARQL Queries in Semantic Web Applications using PHP (Extended Version)

Fatmah Bamashmoos[1,2], Ian Holyer[1], Theo Tryfonas[1], Przemyslaw Woznowski[1]
1 Faculty of Engineering, University of Bristol, Bristol, United Kingdom
2 Faculty of Computing and Information Technology, King Abdulaziz University, Jeddah, Saudi Arabia

*Abstract*— **The Semantic Web (SW) is a significant advancement in the field of Internet technologies and an uncharted territory as far as security is concerned. In this paper we investigate and assess the impact of known attacks of SPARQL/SPARUL injections on Semantic Web applications developed in PHP. We highlight future challenges of developing robust Semantic Web applications using PHP. Our results demonstrate and quantify impacts on Confidentiality, Integrity and Availability (CIA) breaches of data in Semantic Web applications. Our recommendations are targeted to PHP developers, to encourage them to integrate security as early in their design and coding practice as possible.**

## I. Introduction

Throughout the history of Internet technology, from Web 1.0 to Web 2.0 and Web 3.0, significant security issues have arisen. As the Semantic Web (SW) is a recent innovation of the Internet world, and it imports different data from different applications and resources, the possibilities of security issues increase. Masses of people, companies, universities and governments use the Internet. Therefore, significant and sensitive data becomes target of cyber-attacks.

Several studies have discussed the security of the SW in different layers. SPARQL, Blind SPARQL and SPARUL injections become known attacks in the SW world. In this paper we cover these types of attacks that challenge the security of SW applications.

Only a limited number of approaches that have studied the SPARQL injection attacks [3][4][14] exist. They have all applied their tests using Java and placed RDF data and the ontology on the Jena framework server. Their work did not (1) use a PHP as development language, (2) use a Sesame as RDF data store and SPARQL engine, (3) demonstrate a risk assessment on the security framework, (4) list all possible solutions or provide any algorithm for mitigation and test the system after mitigation. The research presented in this paper considers all of these limitations. Other research efforts [1] [10] have just touched on the SPARQL vulnerabilities and possible solutions.

On the other hand, several systems have started moving their data to be linked on the world of linked data, starting from building standard ontologies such as a health care system towards smart hospitals for smart cities [35]. With this paradigm shift, and in order to implement a secure SW system, ensuring which web application development languages are ready to be employed before being stuck in the middle, is critical. That is why in this work we attempt to use different languages and taking healthcare as an example.

Our assumption is that there exists a vulnerability on the SW application that is developed by PHP which allow SPARQL/SPARUL injection attacks to break its security with a high risk on the local RDF/OWL and external data. To assess the SW system under such SPARQL injections attacks and the related risks, we implement a sample PHP SW application, apply risk analysis on the system to measure the risks and then try to mitigate and patch the vulnerability for defending purposes.

We performed SPARQL/SPARUL injection attacks on the linked data in the boundary of a particular application and outside it. In addition, we found that there is no such tool in PHP to mitigate these attacks comparing with Java language. As a result, we provide a filter algorithm to prevent such attacks and provide recommendations for PHP developers toward secure SW application using PHP.

This paper has been organised as follows. Section 2 discusses related work, section 3 describes the implementation of a SW application taking healthcare as an example, section 4 demonstrates relevant attacks by providing different goals for attacking the SW applications, section 5 presents performing the attacks against the example Healthcare SW System, section 6 demonstrates the results of the experiment in addition to the analysis of the problem and the risk assessment, section 7 shows different possible solutions for preventing the attacks in addition to providing a filter algorithm to solve the problem and finally section 8 is evaluation.

## II. Related Works

In the world of massive information, contribution, collaboration, education, business and trade marketing through the Internet, security has become a challenge that has to be studied in order to provide secure places for information storage and exchange. It is always said "Security is not a subject and Security is a verb", and to achieve the goal of the security, three aspects should be studied: Confidentiality, Integrity and Availability known as CIA [11].

The most essential thing on the web is information. However, the purpose is not just to have big data for the sake of it, but to have better, trustful and secure data. As a result, any improvements in that domain should achieve a better outcome for users to feel empowered by it. When users collaborate, they want to keep their privacy and thus can trust when they share their sensitive information. The secrecy of this information depends on the Confidentiality, Integrity and Availability [11].

However, the web and web applications are a target for cyber-attackers. In order, to achieve the goal of being secure in the SW era, security researchers have begun to study the security of SW technology from different aspects and in different layers [12][13][14][16][2][8][9].

One of the most common breaches of security in web 1.0 and web 2.0 is SQL injections. Malicious SQL statements are injected into an entry field of a form on the website. SQL injection takes advantage of security weaknesses in the application's software. The input by the user can get access to specific information or it can delete, corrupt or transfer the entire database [20].

In addition to the common attacks of SQL Injection, Light Weight Directory Access Protocol (LDAP) Injection and XPath Injection, are new techniques that can compromise the new mechanisms [5]. Main ontology query language libraries do not provide any way to avoid the new code injections of a SPARQL Injection, Blind SPARQL Injection and SPARUL Injection [3].

Early work [3] has identified the SPARQL/SPARUL injection, and applied their attack on a SW application implemented in Java, and stored the data in Jena apache server. Researchers [4] have tested their prototype system under SPARQL injection, analyzing the attack. Additionally, recent work (in 2016) [14] has built an insecure SW application to be an environment to allow researchers, students, developers to understand the attack by applying it on their system.

## III. Healthcare Semantic Web System Implementation

The healthcare SW System is an application that has been developed in order to implement attacks, taking an example healthcare system as a case-study because of the importance of the security and privacy of medical data. This development is described briefly in this section.

### *A. Why Healthcare*

Healthcare data is sensitive and must remain private and secure. Work in [11] has started to build a healthcare ontology toward a smart hospital for smart cities. Consequently, this work can give some recommendations for the healthcare system developers to consider in their applications.

A Healthcare Semantic Web System (HCSWS) is a Semantic Web application that is partly developed with the intention to apply SPARQL injection attacks to examine their impact on a particular data server and a particular language.

### *B. Requirements Analysis*

The implementation purpose is to investigate the SW vulnerabilities. The implementation captures doctors' information, patients' information and doctor reports. The data has been chosen randomly with regards to having some critical and sensitive information typical for healthcare systems to build a linked data.

Fig. 1 is illustrates the processes applicable for the HCSWS, and is just a small part of a complete healthcare system where typical actors are Doctors, Patients and Nurses. For brevity this work focuses on the Nurse because this is enough to carry out the intended experiment:

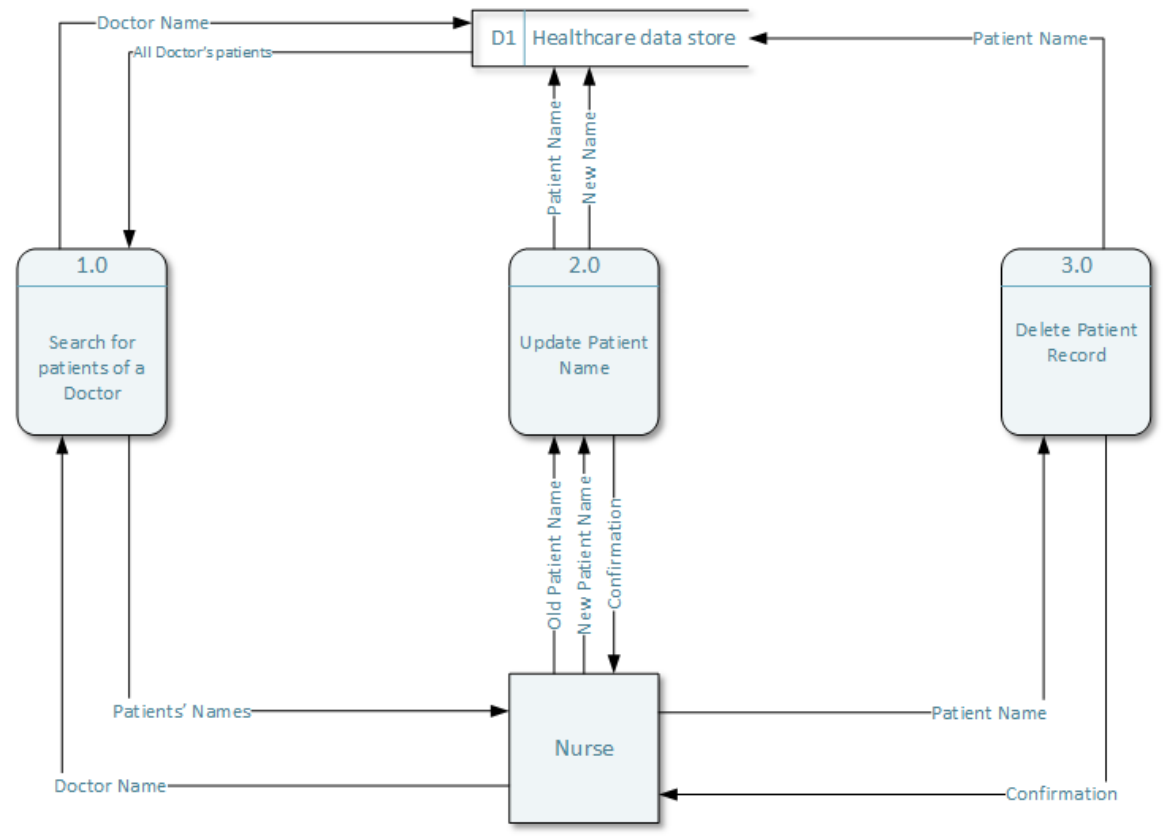

Figure 1. The Data Flow Diagram of the Healthcare Semantic Web system

The data flow diagram shows that there are three processes in the HCSWS: search for patients' names of a particular doctor, update a particular patient's name and delete a particular patient's record. Each one of these processes is divided into minor processes. Process 1.0 is divided as four minor processes: Receive the doctor name from the user, Send the name to the healthcare data store, Check whether the name exists and Return the results. Process 2.0 is made up of four minor processes: Receive the old and the new name from the user, Send the name of a patient, Check whether the name exists, Send the new name and Update (Replace) the old name with the new name. Process 3.0 is divided into four minor processes: Receive the patient name from the user, Send the name of a patient, Check whether the name exists and Delete the triples of that patient.

### *C. HCSWS Design & Implementation*

The HCSWS designed and implemented using PHP 5.5.12 running on the WAMP server, the data implemented using RDF turtle and stored in Sesame 2.8.6 store server. Fig. 2 represents the architecture of the HCSWS. For the RDF data query, SPARQL 1.1 used. EasyRDF library used for communicating

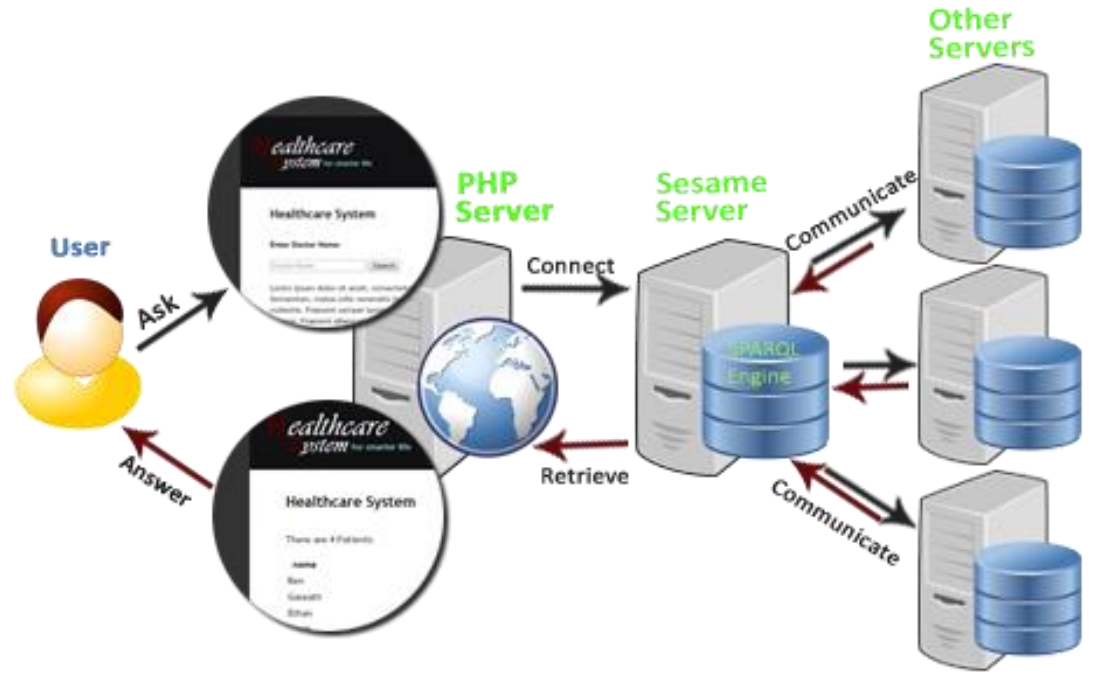

Figure 2. Architecture of the HCSWS

PHP with SPARQL engine.

## IV. Attack Experimental Design

This section describes the design of the experiments of SPARQL injections attacks on the HCSWS. It also provides definitions and goals for each attack.

We implemented different malicious codes to examine the system under these attacks. In addition, we target various healthcare data, as being valuable on the HCSWS in order to assess the risk of the attacks and to check their effect on the Confidentiality, Integrity and Availability of the HCSWS.

Let us consider the following scenario: the nurse exploits her authority for accessing some data to access something she is not supposed to have access to. In other words, the nurse will act as an attacker and more formally, the threat agent will be a malicious nurse.

From the previous section, the intended goal of each process of the HCSWS is clear in addition to the results from each one. However, We can see how the malicious nurse can have another goal to achieve by injecting the user input with a malicious code.

**Definition 1:** Injection attack is a threat on a vulnerable user input by adding malicious code after a required input. This code follows SPARQL Syntax to be combined with the actual query that asks for user input.

## *A. SPARQL Injection*

The attacker in this injection targets the user input in the search screen in Fig. 3. As we have seen, the required input of the search input is a doctor name, and the target goal is to retrieve a list of that doctor's patients' names. Having said that, the nurse had another goal. Did she succeed?

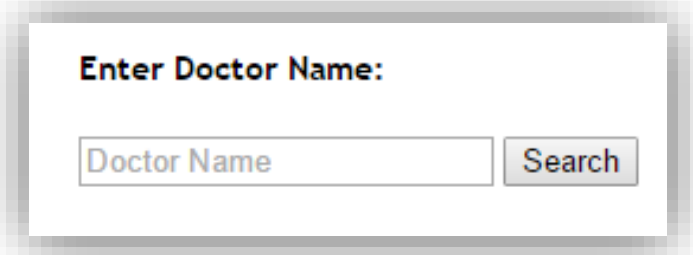

Figure 3. Part of the Search Screen in the HCSWS

For the experimental design, the doctor name has been typed, followed by injection code, which is a SPARQL query. In this design, we have assumed that the attacker has guessed the correct variable for select SPARQL query.

**Definition 2:** A SPARQL injection is an injection attack in which the attackers can guess the select variable to ask for whatever they want.

**First malicious goal:** The nurse's target is to read a particular patient's medical reports instead of patient's name. Technically, her target is to read one of the local RDF data items on the HCSWS that is not supposed to be accessible to her.

**Malicious code design:**

```
Mark".
?p foaf:firstName "Sarah".
?m hc:reportFor ?p.
?m hc:reportDescription ?name. }#
```

*INPUT 1: MALICIOUS CODE 1*

**Note:** for readability, the code is separated from text on a gray background

**Second malicious goal:** The nurse's target is to know all HCSWS data type. Formally, her target is to read all local ontologies on the HCSWS.

**Malicious code design:**

```
Sam".
?a ?name ?b.
}#
```

*INPUT 2: MALICIOUS CODE 2*

**Third malicious goal:** The nurse's target is to know all names in dbpedia. In other words, her target is to read global RDF on the dbpedia, which has this URI < http://DBpedia.org/sparql>.

**Malicious code design:**

```
Sam".
SERVICE <http://DBpedia.org/sparql>
{
SELECT ?name
WHERE{ ?a foaf:name ?name.} LIMIT 50}}#
```

*INPUT 3: MALICIOUS CODE 3*

**Note:** Dbpedia is a SPARQL engine for querying sophisticated data from Wikipedia. The goal is to check the possibility of accessing remote data through the HCSWS.

**Fourth malicious goal:** The nurse's target is to know all properties types that have been used in dbpedia. More formally, her target is reading global ontologies on the dbpedia.

**Malicious code design:**

```
Sam".
SERVICE <http://DBpedia.org/sparql>
{
SELECT DISTINCT ?name
WHERE{ ?a ?name ?b.} LIMIT 50}}#
```

*INPUT 4: MALICIOUS CODE 4*

## *B. Blind SPARQL Injection*

Likewise, the target in this injection is the user input in the search screen “Fig. 4”. The nurse will have different goals to achieve, instead of the intended one. Indeed, it is similar to SPARQL injection, even on the design with a small difference. Assuming that the attacker is unaware of the select variable. In other words, she/he cannot guess the variable name that has been used in the actual query.

As it has been done in the SPARQL injection, by inserting malicious code after the doctor name, the same can be done to perform Blind SPARQL injection.

**Definition 3:** A Blind SPARQL injection is an injection attack in which the attackers cannot guess the SELECT variable. Even though, they are trying to track their malicious goal by the way of asking queries one by one to get a true or false answer.

**First malicious goal:** The nurse's target is to know all patients' emails. More formally, her target is to read some unauthorised local RDF data on the HCSWS.

**Malicious code design:**

```
Sam".
?a hc:editedBy ?b.
?a hc:reportFor ?c.
?c foaf:firstName ?d.
?d foaf:email ?n.
FILTER regex(?n,"^B*")}#
```

*INPUT 5: MALICIOUS CODE 5*

**Second malicious goal:** The nurse's target is to know the reportDate data type. Stated differently, her target is a particular local ontology on the HCSWS.

**Malicious code design:**

```
Sam".
?a ?n ?b.
FILTER regex(?n,"^H*")}#
```

*INPUT 6: MALICIOUS CODE 6*

**Third malicious goal:** The nurse's target is to know the occupation of the person whose name is Thomas in dbpedia. In other word, her target is one of the global RDF on the dbpedia, which has this URI < http://DBpedia.org/sparql>.

**Malicious code design:**

```
 Sam".
SERVICE <http://DBpedia.org/sparql>
{
 SELECT ?n
 WHERE{
   ?a foaf:name "Thomas B. Fitzpatrick".
  ?a dbo:occupation ?n.
   FILTER regex(?n,"^[a-g]*")}}}#
```

*INPUT 7: MALICIOUS CODE 7*

**Fourth malicious goal:** The nurse's target is to know the data type nationality to check if it is used or not. More formally, her target is to know one of the global ontologies on the dbpedia.

**Malicious code design:**

```
Sam".
SERVICE <http://DBpedia.org/sparql>
{
SELECT ?n
WHERE{ ?a ?n ?b.
  FILTER regex(?n,"^na*")
}}}#
```

*INPUT 8: MALICIOUS CODE 8*

## C. SPARUL Injection

The targeted screens in this attack are the update and delete screens as it presented in Fig. 4 and Fig.5. The nurse here, as before, had different malicious goals.

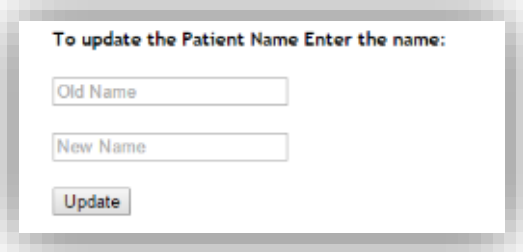

**Figure 4.** Part of the update screen in the HCSWS

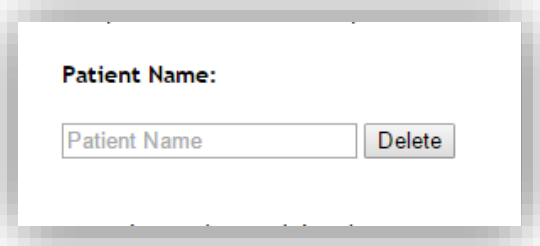

**Figure 5.** Part of the Delete screen in the HCSWS

**Definition 4:** A SPARUL injection is an injection attack that deals with the SPARQL update function to apply a malicious goal by a malicious user. This injection might use the delete or insert technique to corrupt the data.

Firstly, looking at the update screen, the required inputs are the old name of a particular patient and the new name to be updated. The attack target input in here is the new name input.

To apply the attack, the old name will be written, in addition to the new name followed by malicious code. The attacker does not have to know any SPARQL variable, which makes it easier.

**First malicious goal:** The nurse's target is the new name input to add medical report for that patient. Technically, her target is adding data on the local RDF on the HCSWS.

**Malicious code design:**

```
Ethan";
hc:medicalCondition hc:R7.
hc:R7 hc:reportDescription "Lorem ipsum dolor
 sit amet, cueir   mod contentiones nam, his
 no aliquam
WHERE {
?a ?b ?c.
}#
```

*INPUT 9: MALICIOUS CODE 9*

Another formula without #

```
Ethan";
hc:medicalCondition       hc:R7.
hc:R7 hc:reportDescription "Lorem ipsum
 dolor sit amet, cu eirmod contentiones nam
 ,";
```

*INPUT 10: MALICIOUS CODE 10*

**Second malicious goal:** The nurse's target is the new name input to add a mental health new field in the patients graph in order to add some data. In other words, her target is adding a new property to the ontology.

**Malicious code design:**

```
Ethan";
hc:mentalHealth "Lorem ipsum dolor sit
 amet,cu eirmod contentiones nam,".}
WHERE {
  ?a ?b ?c.
}#";
```

*INPUT 11: MALICIOUS CODE 11*

Secondly, in the delete screen shown on Fig. 5, the required input in the delete screen is a particular patient name and the objectives is to delete all of this patient's records. Nevertheless, the malicious nurse had different goal to achieve.

**First malicious goal:** In this scenario, the malicious nurse target is the patient name input to delete all data on the Healthcare system. A bit more formally, her target is deleting all subjects, predicates and objects on the local RDF on the HCSWS.

**Malicious code design:**

```
Gareath".
?a ?b ?c.}
WHERE{
?a ?b ?c.
}#
```

*INPUT 12: MALICIOUS CODE 12*

**Second malicious goal:** The attacker target is the patient name input to delete all property fields. In other words, her target is to delete all the relations and predicates on the HCSWS.

**Malicious code design:**

```
Gareath".
?a ?c ?b.}
WHERE{
?a ?c ?b.
}#
```

*INPUT 13: MALICIOUS CODE 13*

# V. RUNNING THE EXPERIMENTS

This section presents how to run the experimental attacks on the HCSWS.

## A. The HCSWS versus the Attack

To run our attack experiment on the HCSWS, we will inject the designed malicious codes to the target input.

The search input in the HCSWS was the SPARQL injection target and the Blind SPARQL injection target. Therefore, the search input has been injected by their designed malicious codes to be proceed by code1. For example, injecting input 1 to search input.

```
<?
1   $name= $_POST['name'];
2   $query= "
3     SELECT DISTINCT ?name
4     WHERE {?s foaf:firstName \"" .
5     $name . "\". ?r hc:editedBy ?s.
      ?r hc:reportFor ?p.
      ?p foaf:firstName ?name.}
6     ";
7   $result = $sparql->query($query);
?>
```

*CODE 1: THE CODE OF THE SEARCH INPUT TREATMENT ON HCSWS*

**Note:** line 5 is split into two lines for display purposes.

Furthermore, the new user input, which was the target for a SPARUL malicious user, had been injected by (input 9 and 11) where it was treated by the update code to perform the change as shown in code 2.

```
<?
 $old_name= $_POST['old_name'];
 $new_name= $_POST['new_name'];
 $query= "
   DELETE {
     ?p foaf:firstName \"" . $old_name .
"\".
   }
   INSERT {
     hc:P2 foaf:firstName \"" . $new_name .
       "\".} WHERE {?p         foaf:firstName
       \"Gareath\".}
   ";
  $result = $sparql->update($query);
?>
```

*CODE 2: THE UPDATE TREATMENT ON THE HCSWS*

Ultimately, the target for some of the SPARUL injection attacks was the user input on the delete screen. It was injected by input 12 and 13. The input was posted to be processed by code 3 which is the code for processing delete function by PHP.

```
<?
 $name= $_POST['name'];
 $query= "
   DELETE {
     ?p foaf:firstName " . $name . ".}
      WHERE{?p foaf:firstName \"Ethan\".}
 ";
 $result = $sparql->update($query);
?>
```

*CODE 3: THE PHP CODE OF SPARQL DELETE TREATMENT ON HCSWS*

By testing the HCSWS under the injections attacks, our experiment was completed.

## VI. RESULTS AND ANALYSIS

Through the experiments, it has been shown that the SPARQL/SPARUL injection attacks represent high risks. This section discuss the results of our injection attacks, with a critical analysis of these attacks in addition to a risk assessment.

### A. *Experimental Results and Analysis*

After applying the attacks on the HCSWS, it was found that all malicious goals had succeeded, and the attacker achieved what they intended. The attacks targeted the local RDF, OWL and the external RDF, OWL. The following illustration of the results in "Table 1" demonstrates the impact on the local and external RDF data and ontologies.

TABLE 1: THE RSULTS OF THE SPARQL/SPARUL INJECTION ON THE HCSWS AND ON THE EXTERNAL RDF AND OWL

| Assets / Injections | Local RDF | External RDF | Local OWL | External OWL |
|---|---|---|---|---|
| **SPARQL** | Read | Read | Read | Read |
| **Blind SPARQL** | Read | Read | Read | Read |
| **SPARUL** | Write \| Delete | —— | Write \| Delete | —— |

These injections are threats on the CIA of the SW. We can infer that, with SPARQL and Blind SPARQL injection attacks, confidentiality of the data has been lost. Meanwhile, the authenticity and integrity of the local data has been compromised by the SPARUL injection in addition to data availability. That is, the security framework has been affected by these injections. Table 2 highlights the potential impact of these attacks on the CIA of the SW.

Looking back at the designed SPARQL injection, it is noticeable that the SPARQL variable ?name had been used in the (code 1) for treating the search input and in the (input 1) in the injection design. This means that the attackers have to guess this variable to achieve their goal. Nevertheless, it is not so difficult to guess this variable since it might either be printed from the system as a property field's name or it can be a reasonable variable name.

TABLE 2: THE RISK IMPACT ON CIA OF THE SEMANTIC WEB

| Security Objectives | Risk Impact Example |
|---|---|
| **Confidentiality** | Unauthorized users can gain access to confidential and secret data. An example was given when the malicious nurse was able to access medical reports instead of patients' names.<br>This was illustrated in input 1 and code 1 |
| **Integrity** | The authenticity of the patient reports when the malicious nurse has added a report for a particular patient. Consequently, we cannot trust these reports.<br>This was illustrated in input 9 and code 2 |
| **Availability** | When everything on the system has been deleted, the data would not be available and the system would not work correctly.<br>This was illustrated in input 12 and code 3 |

The Blind SPARQL injection attacker had blindly tried to ask the query function recursively about their goals. The query function replied by nothing to be printed and there are no mistakes when it is right. Conversely, if the regular expression is not on the domain of what the attacker asked, the query function gives an error message. As a result, the attacker does not need to guess any variable. However, it can take some time to get the result. Although, the attacker can use some techniques to decrease the time by using letters range in the regular expression in the code. So instead of using ^B* in (input 5) for example, we can use ^[A-M]* which will decrease the time of asking. On the other hand, it might be not logical to use this injection to target the predicate since it is always a URI which is too long to use the blind way to know it.

The SPARUL injection had used a different function to be achieved, which is the update function instead of the query function. Therefore, the attackers cannot exploit the query input for deleting or adding, they have to use the update input for this purpose. However, in our experiments, the malicious nurse used the update screen to add some malicious code to be performed by the update function.

It was found that the easiest and most dangerous attack is the SPARUL injection if there is a chance of using the update function. The attackers does not need to guess any variable or test the query; they just have to follow the syntax of inserting or deleting. Moreover, it can be seen in (input 10) that the attack can be successful even without using a hash sign if it was in a particular format.

Having said that, there are reasons to allow these attacks to occur. To understand what they are, we have analysed one of the SPARQL code example that has received the injection in order to find out the reasons for these attacks and try to find the weaknesses.

By analysing the search processing code (Code 1) for the query function as it presented in Fig. 6, it was found that the structure of the SPARQL query is divided into SPARQL reserved words, SPARQL variables, relations from ontologies that are represented by URIs and SPARQL punctuation marks in addition to the PHP variable.

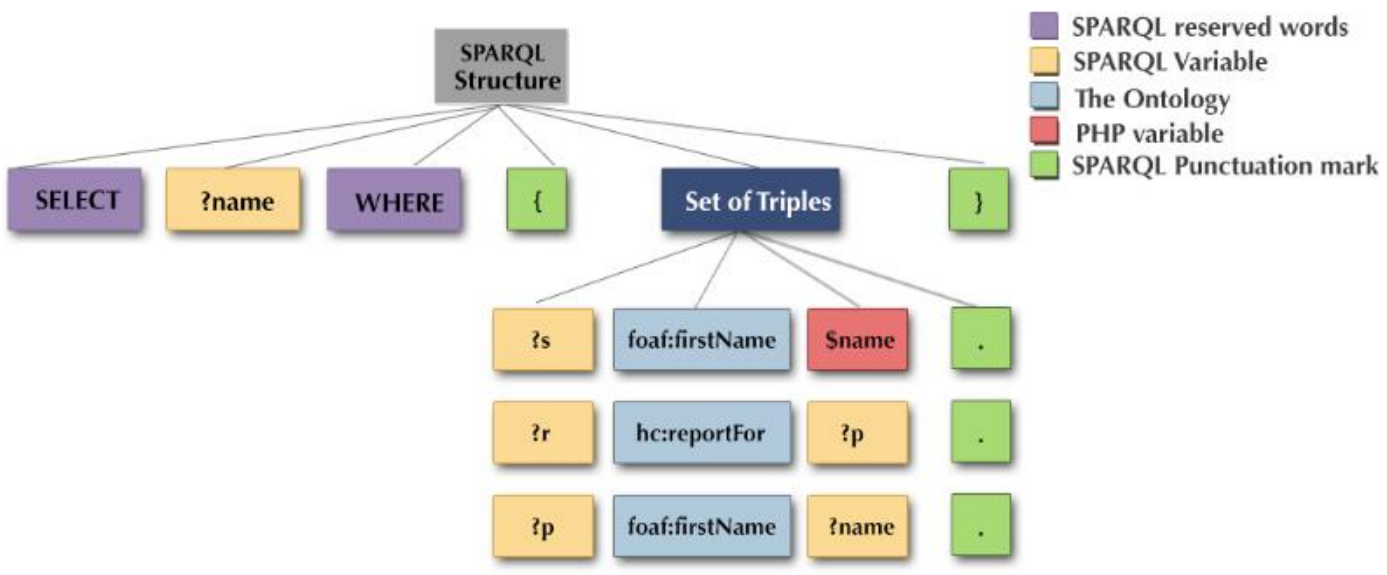

**Figure 6.** Analyzing the content of SPARQL code inside PHP

From the analysis, the weak point is the PHP variable that is supposed to receive a user input from the user without checking what the input might be. Is it the required input or not? If it is not the required, what might it be?

To understand the mechanism of the attack, let us look back at our experiment. For instance, when the malicious nurse used (input 1) to inject the search input on the HCSWS, the search input has sent the input content to (code 1). The PHP variable $name in (code 1) has been exchanged by the input content then it was combined with the code. After that, the SPARQL query function processed the code by SPARQL query syntax. As the attacker had ended their malicious input by hash sign, which is a comment command in SPARQL syntax, everything after the hash sign on the same line had been commented out. The final SPARQL query that was sent to the SPARQL engine on the Sesame server was the following:

```
SELECT DISTINCT ?name
WHERE {?s foaf:firstName "Sam".
  ?p foaf:firstName "Ben".
  ?m hc:reportFor ?p.
  ?m hc:reportDescription ?name. }
```

OUTPUT 1: THE SPARQL CODE AFTER QUERY FUNCTION

Consequently, the malicious goal has been achieved instead of the HCSWS target goal and the attacker had changed the path of the RDF graph.

It was found that the attacks succeeded when they meet the following conditions:

**From the developer's side (HCSWS):**

- The user input was not validated.
- The code was in a particular format, which accept the attack (in one line).

**From the attacker's side (attack experiment design):**

- The attacker followed the required input by a SPARQL code.
- The attacker attached a hash sign to the end of the code to comment out all of the following on the actual code.

## B. *Risk Assessment*

The most valuable content in the HCSWS is the data that has to be protected, so the asset is the local RDF data and the local OWL in addition to the external one. Stating differently, in the real world, we have to protect the money from being stolen. Likewise, the most important thing in the linked data world is the data itself, we have to protect the information especially the sensitive one.

After analysing the code, it can be understood that the threat on the HCSWS is when the threat agents can take advantage of writing whatever they want on the user input without validation. As a result, the threat agent on the HCSWS was the malicious nurse and the threat is the malicious code. The user input is a system vulnerability in this situation. Risk analysis is presented in Fig. 7.

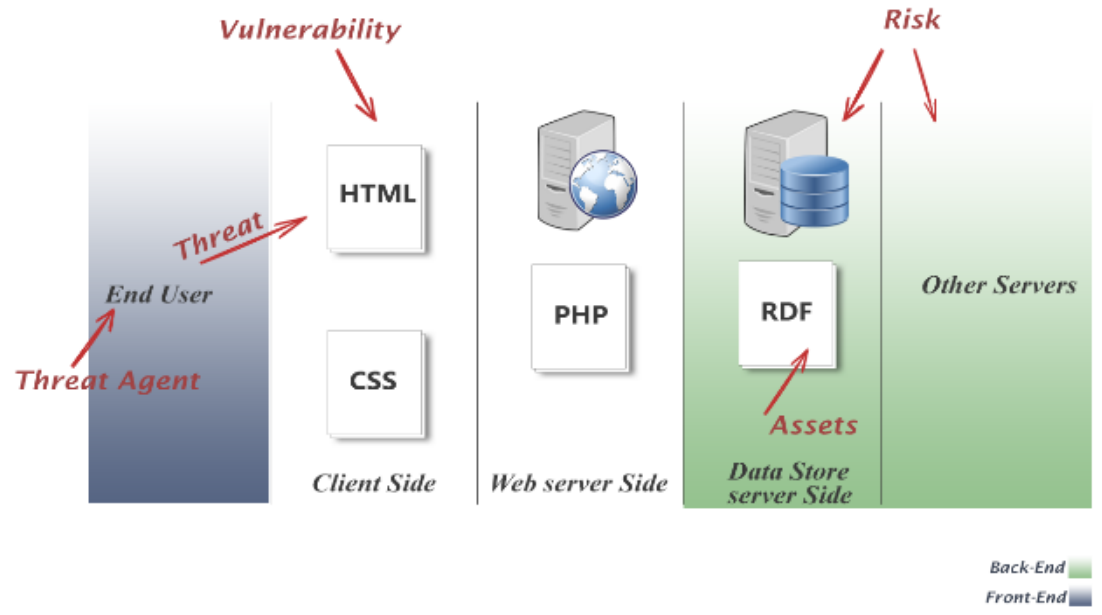

**Figure 7.** The Risk analysis of the SPARQL/SPARUL injection on the HCSWS

From the attacks that were successful, we notice that the malicious code has enforced the SPARQL code to have a different path on the HCSWS graph instead of the correct one. Meanwhile, to escape the target goal by commenting it out and targeting another one. The attacker can either change the path or start another path on the graph. SPARQL is an easy language to query the RDF data. Having the start point, you can continue to the end, crawling from the start node by using subject, predicate and object until the goal is reached. All data properties and types are from standard ontologies. Thus, everything is accessible.

For example, in the experiment the attacker has decided to start a new path. Taking input 1 that has injected the code 1 for example, the target goal was all patients' names for a particular doctor. However, the attacker decided to have description record for the patient Sarah instead. Therefore, she has started with the information that she had, that the patient name is Sarah, asking for her medical report and then her reports description. Hence, she jumped from the main graph to start from another one by specifying the start point, which was the patient Sarah, instead of continuing after the name of the doctor by using semicolon as the SPARQL syntax demonstrated in Fig.8.

As a result, the extension of the attack might reach remote data in addition to local data. The attacker can target any RDF graph just by using subject, predicate and object way. This ability to access any local or external data makes us aware of the powerfulness of the SPARQL injections abilities and this ability from the powerfulness of the SPARQL query language. Still, the injection can be prevented with some effort. More on the risk mitigation in the following section.

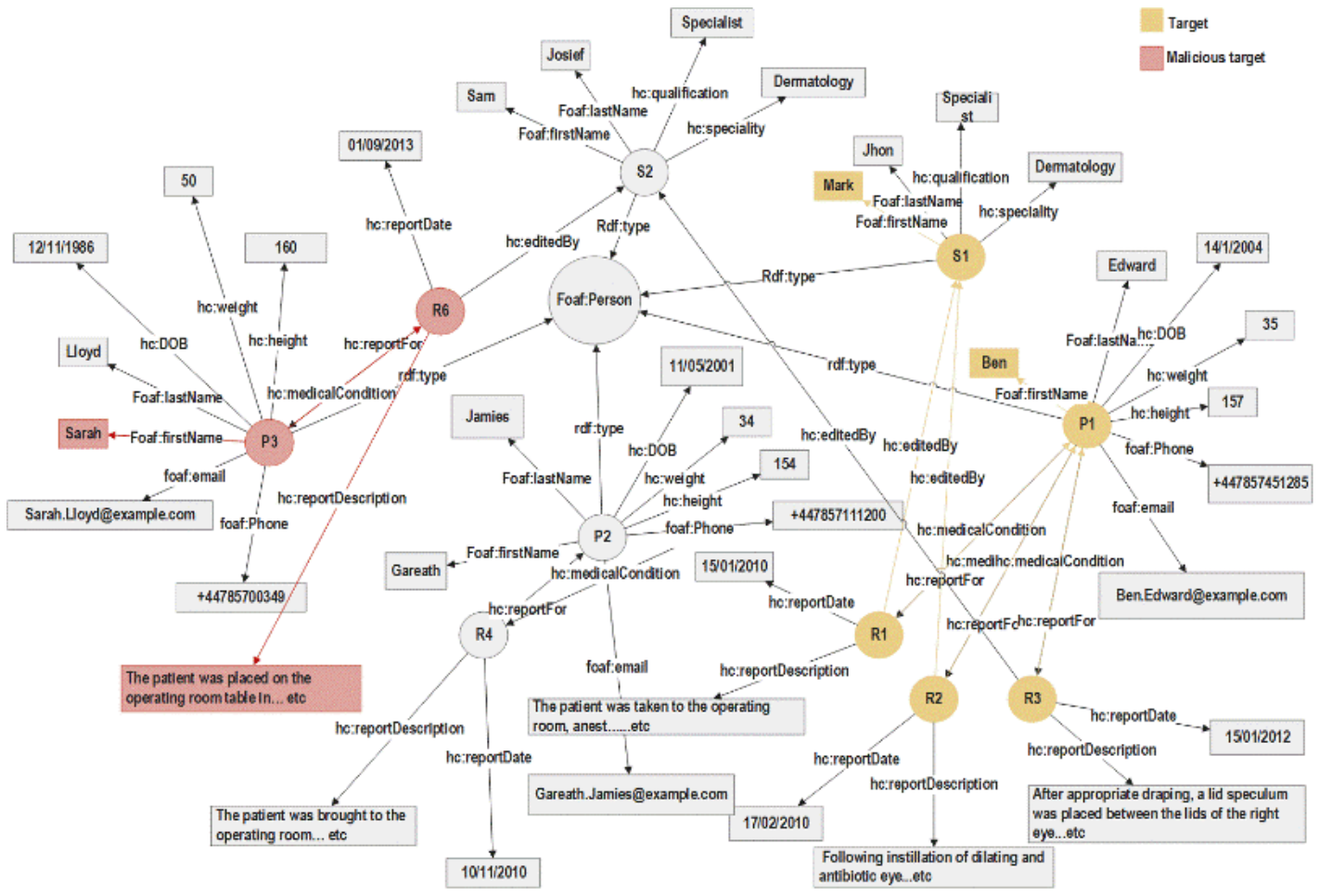

**Figure 8.** The presentation for how the SPARQL injection change the path of the target goal to malicious goal using input 1

## VII. Countermeasures and Risk Mitigation

We have shown that SPARQL/SPARUL injection attacks can have significant impact on the HCSWS, as it was unprotected from these injections. This applies to any SW application that has the same vulnerability. In this section, different types of safeguards have been tested in the HCSWS. These countermeasures are always recommended to prevent any injection attacks similar to SQL injection. Our recommendations are to: use ParametrizedString, define permissions and assign access control for each user and filter and validate the user input.

### *A. ParametrizedString Tool*

In previous studies [6], the prepared statements have been successfully used to prevent SQL injection. Following studies [13][1] have recommended to use the same techniques by providing ParametrizedString to protect the system from the various SPARQL injection types. As a consequence, Jena API [7] has provided this tool for the Java developers to mitigate the vulnerability of SPARQL/SPARUL injections attacks. On the other hand, there is no such tool to be used in PHP. Therefore, we could not attempt to apply this countermeasure.

### *B. Permission and Access Control*

Assigning permission and access control for the system user is one of the suggested solutions for preventing attacks. However, in our experiment on the HCSWS, the authorized user who is the nurse, is the one who decided to exploit this permission to add malicious code. Even though, it is important to assign privileges for each user on the system, not all authorized users are trustful. Therefore, the mitigation should not depend on the users; but rather, it should have a powerful security by itself. There are no reasons to leave the vulnerability without mitigation by specifying the permissions since this will not satisfy the security.

### *C. Filtering User Input*

In the previous section, the query code in the HCSWS that processed the user input has been analyzed to understand the reasons of the attacks and to identify the threat and the vulnerability in order to mitigate it. As it was found that the user input is a major problem of the injection, filtering is one of the suggested solutions.

Filtering is a technique to validate the user input from any unexpected input to be posted to the system. Our filtering method is to try to reject any punctuation marks and any reserved words in SPARQL query language that are not expected as an input to our system. We have provided a filtering algorithm to be applied by using any languages to filter the user input from SPARQL injection attack (Algorithm 1).

After applying the filter algorithm by using PHP, the HCSWS has been tested against SPARQL injections attacks again to check the HCSWS's defence. Consequently, the injection has failed and the defence has succeeded.

The mechanism of the filtering method is to validate any input from having any illegal content for a particular input. Thus, each input has a different purpose and different content expectation and the developers should be aware of this issue to apply the algorithm correctly.

```
FUNCTION FILTER (UserInput)
  DEFINE ARRAY of SPARQL reserved words and
  all possible
  Punctuation marks that we do not expect from
  the user
  FOR i UPTO ArrayLength
    COMPARE UserInput with
    ARRAY[i]
    If COMPARE RETURN Equal THEN
      RETURN TRUE
     END
  END
  RETURN FALSE
END FUNCTION
```

*Algorithm 1: Filtering the User Input from SPARQL Injections Attacks*

With different situations of each input, there is a probability that the user input might require to have SPARQL code. In this situation, the developer has to ensure that the sub query code does not have the same PHP variable. The reason is that if the content has the same variable, the whole result would change since the inner variable will discard the outer one by its new contents. Additionally, even if the input would have SPARQL content, the last character of the content should not be a hash sign.

### D. *Client side vs. Web Server side vs. Data Server side*

As SW systems have several components, it is worth to think about in which component the protection against injection attacks should be implemented. Here is a typical scenario: the end user who acts as an attacker deals directly with the client side that is published by a web server. The web server has contact with the data store server to retrieve the required information. The concept of the SW is that the data is stored on a different server and not on the same server of the entire system as before. The question here is which side has to mitigate the vulnerability.

Between the vulnerability and the risk, see Fig. 8, the threat of the injection code should be prevented. The vulnerability is a user input, and the user input is captured via HTML. PHP is the programming language used to act as an intermediary between the RDF data on the Sesame server or on the Jena Fuseki server and the end user. The risk as it was found occurs on the data server side where the RDF and the ontologies are located.

The client side, which is the layout of the system that is implemented in HTML and CSS, can have a technique for validating the user input by using JavaScript language. This can prevent any undesired inputs from being posted to the web server where a PHP script processes the content. Having said that, the user might turn off the JavaScript, Thus, the validation will not work and the prevention will fail.

The WAMP server is where all webpages implemented in PHP are located. If user input validation is done here, it can achieve the security purpose by mitigating the weak point by filtering the contents of that user input, as it was proposed before. In addition, using standard code writing will help to suspend the penetration since one of the reasons of the attack is the format of the code that was on the same line. Moreover, if it is possible to use parametrized string, then it should be used to deal with the query/update code, regardless if it is still not offered in PHP.

In the data store server side, we can secure any sensitive ontologies using hash functions to be protected ontologies.

To summarize our discussion, there are different ways to prevent injection attacks. These ways may protect against and prevent security breaches. Suggested solutions include:

- Provide a SPARQL ParametrizedString tool for PHP developers.
- Validate the user input using Filtering algorithm.
- Have a standard for code writing.

Meanwhile, some helpful ways that contribute to attack prevention:

- Assign permission and access control for each user in the system.
- Protect sensitive ontologies by using hash functions.
- Use unpredictable variables names.

## VIII. Evaluation

### A. *Semantic Web and PHP*

This work has found that there are not many tools to be used by PHP to facilitate communicating with linked data. A sparqllib library [17] has been chosen randomly used for the first time to communicate with the RDF of the HCSWS on the Sesame and Jena server respectively. Then, it was discovered that the library does not work with the update function while we are trying to implement the Search and Delete screen in the HCSWS. The EasyRDF [16] is another library that has been chosen to facilitate the communication with the RDF data server and SPARQL engine for either querying or updating RDF data. Nevertheless, this library does not have any tools to support PHP for mitigating the vulnerability of SPARQL/SPARUL injection attacks.

Despite simple and basic capabilities that can be used to develop a secure SW application in PHP, the awareness of filtering user's inputs can satisfy the security aims.

On the other hand, it was found that Jena API (the library that facilitate the communication between Java and data store server) has provided the ParametrizedString tool to support the java developers to build secure SW application using Java. Thus, there is more facilitation for Java despite the popularity of PHP which might discourage PHP developers to move towards SW. However, our recommendations might guide the PHP developers toward implementing secure SW applications.

### B. *ParametrizedString and Filter method*

The filter method has been successfully used to prevent the injection attack. To use it, the developer has to create an array of all unexpected input and then to compare all array elements with all elements on a particular user input. This way proved to be effective against the SPARQL injection in our experiment. Nevertheless, the developer might find this way tedious and time consuming since each input has to have different contents of array to check that input. Thinking of having one hundred inputs for example, the developer should check each of them against different arrays. Additionally, the developer might feel lazy and might neglect to prepare an array for each input. He/She might think that the reason of the attack is just the comment sign and he/she just has to check the input for having this sign. However, it is possible in some situation, as in SPARUL injection, that the attack succeeded without using a hash sign with a certain format (code 10). Although, the filter is a successful way for preventing injection attack, providing a ParametrizedString tool is highly recommended and encouraged; this supports the recommendation from [4].

The SPARQL ParametrizedString tool works similar to the prepared statement against SQL injection that has proved its ability to prevent the attack [6]. The ParametrizedString would be easier and efficient. The developer has just to use this function to prepare efficiently the query and then connect it with the input variable. So, it deals with the query and the input separately.

Closely related for instance, in the airport checkpoint, the security guard can check the content of the luggage manually. The problem with this is that the security guard might omit or miss something without notice; this is similar to filter method. On the other hand, the security guard can use some tools to check the luggage contents automatically as we can see at the airport checkpoint. The automatic method is more efficient and easier and this is like ParametrizedString tool.

Our argument here is not to say that filter algorithm would not succeed, but to prove that there is a better solution but it is not provided yet. On the other hand, despite the advantages of ParametrizedString, the filter method can be used to validate user input and protect against any injection attacks, whether it is SQL, SPARQL or any other.

Moreover, the ParametrizedString will be used with SPARQL code. On the other hand, the prepared statement is used with SQL code. The point here is different functions should be used with different types of query languages to be protected from the injections, with nearly the same concept, and this lead to question that how we can unite these function to be one for protecting any injections. This question has to be studied in future works.

### C. Vulnerability vs. Responsibility

From the experiments, it was found that the user input can be a SPARQL injection attack threat if it is not validated which support the result of [3][4]. There is no doubt that in every system even on Web 1.0, Web 2.0 or the new generation, the user input would be vulnerable without validation.

Nevertheless, the security risk is different, since the type of the asset and storing the asset is different. From a security point of view, protecting valuable things is critical regardless of how they are stored. However, these valuable things might be different and they should have different security approaches. Meanwhile, when the risk would lead to a true disaster in the information security, there will no place for negligence from the security guard, and that is why the risk assessment has been included. The SW is a world of linked data, and towards linked data, weaknesses gates should be securely closed toward a securely linked world.

On the contrary, it was found from the experiment that the first responsible for the attack is the developer, since the one reason of the attack is the format of the actual code that received the injection as it was seen in results section.

Vulnerability is a responsibility. Responsibility for whoever acts as a developer, security agent, researcher or an ontologist. Responsibility for what? Responsibility for writing the code efficiently while paying attention to security aspects, and not to simplify the injection. Responsibility for standardising the code writing. Responsibility for choosing unpredictable variables names. Responsibility for encrypting sensitive ontologies. Responsibility for validating any user input and responsibility for providing tools to mitigate the vulnerability.

## IX. Conclusion

Several studies have started to evaluate the security of the SW. This research continued to study some vulnerabilities on its applications. These vulnerabilities pose security concerns and might allow SPARQL, Blind SPARQL and SPARUL injection attacks to break the security framework of SW applications.

The research approach was to apply the attack on a sample of SW application that is developed by PHP, taking health care as an example and calling the system HCSWS. The attacks experiments have proved that the system is vulnerable to the SPARQL/SPARUL injections attacks if the user input is not validated or mitigated. In addition, the simple capabilities of using PHP to implement a SW application.

The experiments showed the high security impact on the CIA of the HCSWS. The problem of the attack has been analysed in order to identify risk analysis towards risk mitigation. As a result, the research assumption has been confirmed.

Several solutions have been proposed with some arguments for the best one in addition to Filter algorithm that was provided.

The research has highlighted that vulnerability mitigation is a responsibility of the developer and there is no place for complaisance. The world now is built on information and linked data, so, weaknesses should be patched. Meanwhile, PHP is a popular language for web application development and to cope with the new generation of technologies. PHP should have support for developing secure code to communicate with RDF data stores and retrieving or modifying linked data.

## Acknowledgment

Fatmah is a PhD student who is sponsored by King Abdulaziz University, Saudi Arabia, to study her PhD in University of Bristol. This work is part of her master thesis that was done in University of Bristol. This work was also supported in part under the SPHERE IRC funded by the UK Engineering and Physical Sciences Research Council (EPSRC), Grant EP/K031910/1

## Refrences

[1] K. Sumit, and K. Suresh, "Semantic Web attacks and countermeasures." in *Engineering and Technology Research (ICAETR), 2014 International Conference on*. IEEE, 2014.

[2] Gabillon A. and Letouzey L., A View Based Access Control Model for SPARQL. NSS'10, 2010. pp. 105-112.

[3] O. Pablo, et. Al, "Identifying Security Issues in the Semantic Web: Injection attacks in the Semantic Query Languages." Actas de las {VI} Jornadas Científico-Técnicas en Servicios Web y {SOA}} 51: 4529- 4542.

[4] X. Yang, Y. Chen, W. Zhang and S. Zhang, 'Exploring injection prevention technologies for security-aware distributed collaborative manufacturing on the Semantic Web', *Int J Adv Manuf Technol*, vol. 54, no. 9-12, pp. 1167-1177, 2010.

[5] A. Razzaq, K. Latif, H. Ahmad, A. Hur, Z. Anwar and P. Bloodsworth, "Semantic Security Against Web Application Attacks," *Information Sciences*, vol. 254, pp. 19-38, January 2014.

[6] S. Thomas, L. Williams and T. Xie, 'On automated prepared statement generation to remove SQL injection vulnerabilities', *Information and Software Technology*, vol. 51, no. 3, pp. 589-598, 2009.

[7] Jena.apache.org, 'Apache Jena - Parameterized SPARQL String', 2015. [Online]. Available: https://jena.apache.org/documentation/query/parameterized-sparql-strings.html. [Accessed: 29- Apr- 2015].

[8] A. Medić and A. Golubović, "Making secure Semantic Web." Universal Journal of Computer Science and Engineering Technology 1, Vol. 2, pp. 99-104. 2010.

[9] A.N. Gupta and P.S. Thilagam, "Attacks on Web Services Need To Secure XML on Web." Computer Science & Engineering 3, Vol. 5. 2013.

[10] R. Bruwer and R. Riaan, "Web 3.0: Governance, Risks And Safeguards", *Journal of Applied Business Research (JABR)*, vol. 313, pp. 1037-1056, 2015.

[11] S. Anand and A. Verma, "Development of Ontology for Smart Hospital and Implementation using UML and RDF", *IJCSI International Journal of Computer Science*, vol. 7, no. 5, pp. 206-212, 2010.

[12] B. Thuraisingham, 'Security issues for the semantic web', *2013 IEEE 37th Annual Computer Software and Application Conference.* IEEE Computer Society, pp. 632, 2003.

[13] P. Orduña, A. Almeida, U. Aguilera, X. Laiseca and A. Gómez-Goiri, 'SPARQL/RDQL/SPARUL Injection', *Morelab.deusto.es*, 2010. [Online].

[14] H. Asghar, Z. Anwar and K. Latif, "A deliberately insecure RDF-based Semantic Web application framework for teaching SPARQL/SPARUL injection attacks and defense mechanisms", *Computers & Security*, vol. 58, pp. 63-82, 2016.

[15] M. Farina, "Modern PHP, Popularity, and Facebook | Engineered Web", *Engineeredweb.com*, 2014. [Online]. Available: http://engineeredweb.com/blog/2014/modern-php-facebook/. [Accessed: 08-Jan- 2016].

[16] N. Humfrey, "EasyRdf - RDF Library for PHP", *Easyrdf.org*, 2016. [Online]. Available: http://www.easyrdf.org/. [Accessed: 12- Jan- 2016].

[17] Graphite.ecs.soton.ac.uk, "SPARQL RDF Library for PHP", 2016. [Online]. Available: http://graphite.ecs.soton.ac.uk/sparqllib/. [Accessed: 12-Jan- 2016].